\documentclass[conference,a4paper]{APSIPA2021}

\usepackage{multirow}
\usepackage{graphicx}
\usepackage{amsmath}
\usepackage{amssymb}
\usepackage{amsxtra}
\usepackage{threeparttable}

\usepackage{booktabs}
\usepackage{flushend}
\usepackage[hidelinks]{hyperref}
\usepackage{subfigure}
\begin{document}

\title{Temporal Feedback Convolutional Recurrent \\Neural Networks for Speech Command Recognition}

\author{%
\authorblockN{%
Taejun Kim and Juhan Nam
}
\authorblockA{%
Graduate School of Culture Technology, KAIST, Daejeon, Republic of Korea \\
E-mail: {\tt \{taejun,juhan.nam\}@kaist.ac.kr}
}
}

\maketitle
\thispagestyle{empty}

\begin{abstract}
End-to-end learning models using raw waveforms as input have shown superior performances in many audio recognition tasks. However, most model architectures are based on convolutional neural networks (CNN) which were mainly developed for visual recognition tasks. In this paper, we propose an extension of squeeze-and-excitation networks (SENets) which adds temporal feedback control from the top-layer features to channel-wise feature activations in lower layers using a recurrent module. This is analogous to the adaptive gain control mechanism of outer hair-cell in the human auditory system. We apply the proposed model to speech command recognition and show that it slightly outperforms the SENets and other CNN-based models. We also investigate the details of the performance improvement by conducting failure analysis and visualizing the channel-wise feature scaling induced by the temporal feedback.
\end{abstract}

\section{Introduction}
\label{sec:intro}

End-to-end learning models in the audio domain take raw waveforms as input and learn features from scratch instead of using mel-spectrogram or hand-crafted features. They have shown superior performances in many audio recognition tasks, for example, music auto-tagging \cite{lee2017sample,Pons2018,kim2019comparison}, speech recognition \cite{sainath2015learning, kim2019comparison}, speech emotion recognition \cite{Tzirakis2018,Latif2019} and environmental sound recognition \cite{dai2017very,Abdoli2019}. However, the model architectures are mostly based on CNN, which was originally inspired by human visual processing \cite{fukushima1980} and has been developed for visual recognition tasks in the computer vision community. Considering the temporal nature of audio data, recurrent neural network (RNN) modules are often combined with CNN. However, it is common that the model architecture is heuristically designed. This paper aims to design a convolutional recurrent neural network (CRNN) inspired by human auditory processing.

Neural signal processing in the human auditory system is known to be bidirectional \cite{Lyon:2017}. Cochlear in the inner ears sends signals to the brain via the inner hair cells and the auditory nerve, which is called \emph{afferent} projection. In respond to this, the brain also sends \emph{efferent} signals to the periphery, which terminates on the outer cell hair. The role of efferent signals is to control the gain of the feedforward signals. That is, it adapts to incoming sounds and reduces the stimulus in cochlear with a feedback control. This adaptive gain control has been emulated in computational models of cochlear-filter bank, for example, the Cascade of Asymmetric Resonators with Fast-Acting Compression (CAR-FAC) cochlear model \cite{Lyon:2017}. In the context of neural network, such adaptive gain control mechanism is found in the squeeze-and-excitation (SE) module which scales channel-wise feature responses by explicitly modeling interdependencies between channels in CNN \cite{hu2018senet}. Kim et al. applied the SE module to audio classification tasks and observed that it adaptively controls the channel-wise activations \cite{kim2019comparison}. However, the SE module was connected to the convolutional block in a feedfoward manner.

\begin{figure}[t]
\includegraphics[width=\linewidth,keepaspectratio]{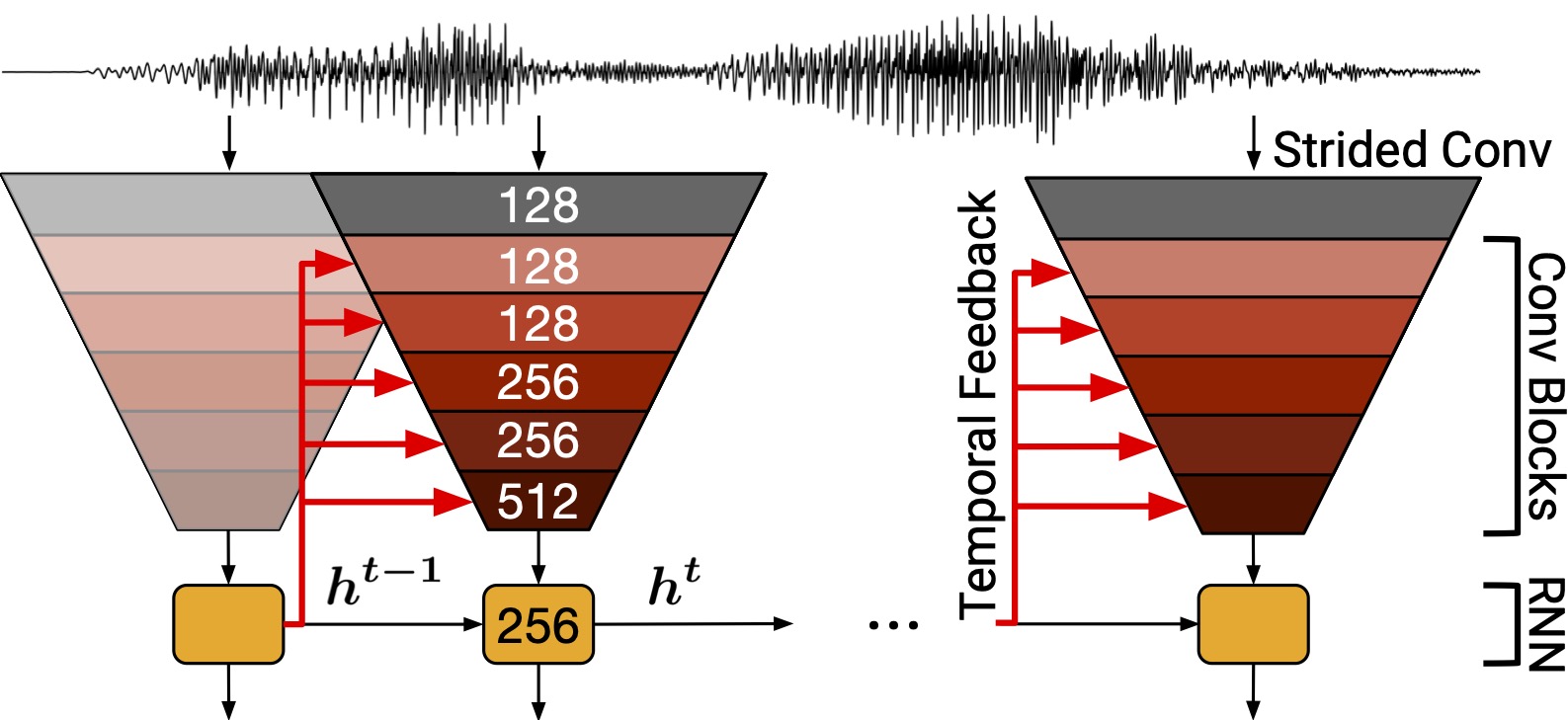}
\caption{
An illustration of the TF-CRNN architecture. The hidden states of the RNN module are used as temporal feedbacks that control the sensitivity of channel-wise feature activations in the CNN blocks. The numbers in each convolutional block and RNN indicate the number of filters and hidden units, respectively.
}
\label{fig:arch}
\end{figure}

In this paper, we extend the SE module by adding the controls of channel-wise feature activations via feedback connections, which is analogous to the efferent control mechanism of outer hair-cell. We implement the module in a CRNN architecture to make the feedback connections causal in time. Specifically, the CNN modules capture local features from the raw waveforms in an end-to-end manner and the RNN module in the top layer controls the strength of channel-wise feature activations in the CNN modules in the next time step via temporal feedback connections, as shown in \figurename~\ref{fig:arch}. We call the model Temporal Feedback CRNNs (TF-CRNNs). We evaluate TF-CRNNs for speech command recognition. The results show that the proposed models slightly outperform the SENets and other CNN-based models on the speech commands dataset~\cite{warden2018speechcommands}. In addition, we investigate the effect of temporal feedback in speech command recognition through failure analysis and visualize channel-wise feature scales in the CNN modules to better understand the operation of the feedback controls in TF-CRNN.

\begin{figure*}[!t]
  \centering
  \subfigure[Basic block~\cite{lee2017sample}]{
    \includegraphics[height=4cm,keepaspectratio]{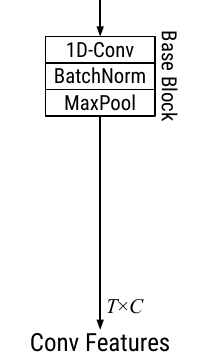}\label{fig:block_basic}
  }
  \subfigure[TF-basic block]{
    \includegraphics[height=4cm,keepaspectratio]{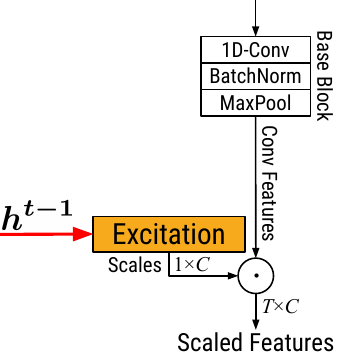}\label{fig:block_tfbasic}
  }
  \hspace{5mm}
  \subfigure[SE block~\cite{hu2018senet}]{
    \includegraphics[height=4cm,keepaspectratio]{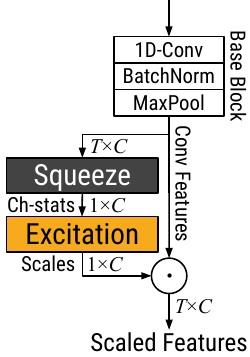}\label{fig:block_se}
  }
  \subfigure[TF-SE block]{
    \includegraphics[height=4cm,keepaspectratio]{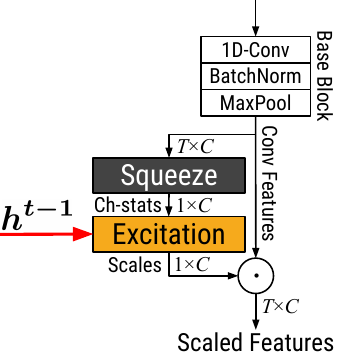}\label{fig:block_tfse}
  }
  \hspace{5mm}
  \subfigure[Res block]{
    \includegraphics[height=4cm,keepaspectratio]{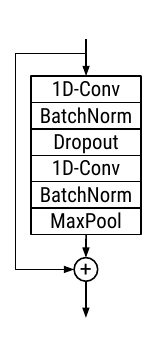}\label{fig:block_res}
  }
  \caption{
    Convolutional building blocks from the previous works (a, c) and the proposed blocks (b, d) based on them.
    The TF blocks take hidden states $h^{t-1}$ to scale each channel of features.
    (e) Res blocks which can be used instead of the base blocks to increase representation power.
    ``Ch-stats" stands for channel-wise statistics.
  }
  \label{fig:block}
\end{figure*}

\section{Temporal Feedback CRNN}
\label{sec:temporal_feedback}

\subsection{Overall Architecture}
The overall architecture of TF-CRNN is based on SampleCNN~\cite{lee2017sample,lee2018samplecnn} and the specific design of the feature scaling in convolutional blocks is based on the SE block.
The SampleCNNs take raw waveforms as input and consists of a strided convolutional layer (stride size of 3) in the bottom layer and 8 basic blocks in \figurename~\ref{fig:block_basic}.
All convolution layers have a kernel size of 3 and a stride of 1 except the bottom layer.
The differences between SampleCNN and TF-CRNN are that the last three convolutional blocks are replaced with an RNN layer in TF-CRNN and the hidden states of the RNN from the previous time step are connected to the convolutional blocks below as shown in \figurename~\ref{fig:arch}.

\subsection{Feature Scaling by Temporal Feedback}
\label{sec:tfblock}
We use the feedback connections from the RNN module at the previous time step to scale each feature map in the convolutional blocks at the current time step. The idea of channel-wise scaling is borrowed from SENets~\cite{hu2018senet}, where the feature map scaling values, the excitation scales, are learned via a separated feedforward block from the convolutional layer. Specifically, the feedforward block, SE block in \figurename~\ref{fig:block_se}, consists of two operations: \textit{squeeze} and \textit{excitation}. The squeeze operation takes global average pooling of feature maps over time. Therefore, $C \times T$ feature maps are reduced to $C \times 1$ channel-wise statistics where $T$ is the input length and $C$ is the number of channels (or filters). The excitation operation then takes the channel-wise statistics as inputs and computes the scaling values with a range of $[0,1]$ through two fully-connected (FC) layers forming a bottleneck. The bottleneck reduces a dimensionality to $C/r$ and then increases it back to $C$ to fit the number of channels, where $r$ is a reduction ratio. In our experiment, we set $r$ to $2^3$ following the previous work~\cite{kim2019comparison}.

\begin{figure}[t]
    \centering
    \includegraphics[width=0.9\linewidth,keepaspectratio]{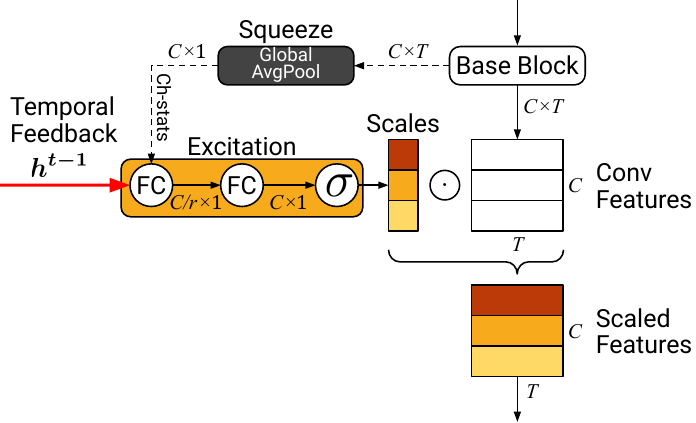}
    \caption{
    A detailed explanation about feature scaling procedure in the proposed TF-basic block (without the dashed line; \figurename~\ref{fig:block_tfbasic}) and TF-SE block (with the dashed line; \figurename~\ref{fig:block_tfse}).
    $C$ and $T$ denote the number of channels and the length of convolutional features in time, respectively. $h^t$ is a hidden state of the RNN at time step $t$. ``FC" stands for a fully-connected layer and $\sigma$ denotes the sigmoid function.
    }
    \label{fig:tfscale}
\vspace{-2mm}
\end{figure}

\figurename~\ref{fig:tfscale} illustrates the process of feature scaling in TF blocks.
The TF-basic block also has the excitation operation as with the SE block, but the TF-basic block takes temporal feedbacks as input instead of channel-wise statistics from the squeeze operation.
In other words, the SE block uses local low-level features to compute the excitation scales, whereas the TF block uses temporally aggregated high-level features.
However, the TF-basic block can be easily extended to the TF-SE block by concatenating the temporal feedback and channel-wise statistics from lower layers as input for the excitation operation.
We also evaluate performances by replacing the base blocks in \figurename~\ref{fig:block}(a-d) with Res block (\figurename~\ref{fig:block_res}) for comparison with the previous work~\cite{kim2019comparison}.
The Res block has doubled convolution layers and a skip connection which helps to stack layers deeply~\cite{he2016identity}.
A dropout layer is inserted between convolution layers to prevent overfitting.

\section{Experimental Setup}
\label{sec:experiments}
\subsection{Dataset}
We used the Speech Commands dataset version 2 for speech command recognition~\cite{warden2018speechcommands}. The dataset contains 84,843, 9,981, and 11,005 audio examples in training, validation, and test splits, respectively. All files have one second in length with a sampling rate of 16 kHz and each of them contains one utterance of commands. The number of the commands is 35 which corresponds to the number of the output classes. We report performances using accuracy (\%) on the test set.

\subsection{Input/Output Setup}
\label{sec:loss}
Many-to-one is a typical input/output setup for RNNs in single-label classification. Due to the vanishing gradient over time, however, it is often difficult to train RNN-based models with a complex configuration. Also, in speech commands recognition, the true commands appear not only at the end but throughout the entire audio sequences. Specifically, timbral characteristics of commands often appear at the beginning of utterances so candidate commands are narrowed down to few commands at the early stage of sequences. For example, when ``f'' comes into the model at the beginning, candidates are narrowed down to 4 commands (``five", ``follow", ``forward", and ``four") among 35 commands. Therefore, the model should be able to make a partial prediction of commands in the middle of speech (We will show the detail of the temporal trajectory of prediction probabilities in Section~\ref{sec:analysis_tf}). With this motivation, we also train the RNN module in the many-to-many setting where the loss is computed every time step and and therefore the gradient flows from each time step of the output. In the test phase, however, we used the prediction only from the last time step. We compare the many-to-many setting to the many-to-one setting in all models.   



\subsection{Implementation Details}
The number of convolutional filters at each block is denoted in \figurename~\ref{fig:arch}. The size of filter and max pooing is fixed to 3. The first convolutional block in the bottom has striding with a size of 3 instead of max-pooling. 
We use Gated Recurrent Units (GRUs)~\cite{chung2014gru} with 256 hidden units to implement the RNN module and initialized the hidden states with zeros.
All neural networks are trained with a batch size of 64 using stochastic gradient descent with Nesterov momentum of 0.9. The initial learning rate is set to be 0.025 and decayed by a factor of 5 when the validation loss does not decrease for 3 epochs.
The training is stopped after three decays of learning rate.
The size of the time step $t$ of CRNNs and TF-CRNNs is set to be 100ms (1600 samples) after a grid search of \{12.5, 25, 50, 100, 250\}ms and the window size (or the input size for the CNN at a single time step) is set to be twice the time step (e.g. 200ms for 100ms time step size).
We inserted a dropout layer with a ratio of 0.5 between the CNN blocks and the RNN layer in all CRNN-based models and before the last layer in all CNN-based models. The dropout ratio in the Res block is set to 0.2.
We used PyTorch~\cite{paszke2017pytorch} for this experiment. The source code is available at the link\footnote{\url{https://github.com/tae-jun/temporal-feedback-crnn}}.

\begin{table}[t]
\centering
\caption{
  Accuracy (\%) for speech command recognition \\on the Speech Commands dataset version 2
}
\label{tab:results}
\begin{tabular}{llcc}
\toprule
Model & Block & Many-to-One & Many-to-Many \\
\midrule
\multirow{3}{*}{CNN~\cite{kim2019comparison}}
& Basic  & 94.63 (0.12) & -- \\
& SE     & 95.33 (0.10) & -- \\
& Res-SE & 95.93 (0.08) & -- \\
\hline \\[-2ex]
\multirow{3}{*}{CRNN}
& Basic  & 94.86 (0.15) & 95.48 (0.21) \\
& SE     & 95.56 (0.12) & 95.94 (0.12) \\
& Res-SE & 95.95 (0.06) & 96.32 (0.22) \\
\hline \\[-2ex]
\multirow{3}{*}{TF-CRNN}
& TF-Basic  & 95.44 (0.02) & 96.03 (0.12) \\
& TF-SE     & 95.51 (0.15) & 96.17 (0.14) \\
& TF-Res-SE & 95.89 (0.22) & \textbf{96.55} (0.11) \\

\hline \hline \\[-2ex]
\multicolumn{2}{l}{HarmonicCNN~\cite{won2020data}} & 96.39 & -- \\
\multicolumn{2}{l}{AttnCRNN~\cite{de2018neural}} & 93.90 & -- \\
\bottomrule
\end{tabular}
\end{table}

\section{Results}
Table~\ref{tab:results} summarizes accuracy (\%) of speech command recognition for all models built with different blocks. All results are averages of 3 runs and the standard deviations of accuracy are denoted in parentheses. The CNN models have the same architecture with the CRNN models except that the RNN layer is replaced with 3 convolutional blocks. CRNN models have the same architecture with the TF-CRNN models except that they have no temporal feedback connections. 
The result shows how each of different settings improves the performance. The comparison between CNN and CRNN models indicates that the RNN layer consistently increases the accuracy for all blocks. In addition, the many-to-many setting in the CRNN models increases the accuracy further. The comparison between CRNN models and TF-CRNN models indicates that the temporal feedback connections consistently increase the accuracy for all blocks in the many-to-many setting. Among them, the best accuracy was obtained from the proposed TF-Res-SE block. In the many-to-one setting, however, the accuracy degrades for all blocks presumably because the vanishing gradient issue becomes more complex with the temporal feedback connections. This can be however addressed by more frequent error back-propagation in the many-to-many setup.

The last two rows in Table~\ref{tab:results} show the accuracy on the speech command dataset from two previous works.
HarmonicCNN~\cite{won2020data} exploits trainable harmonic filters which take spectrograms as inputs and produce inherent harmonic structures of audio signals. The harmonic features are fed into seven 2-D convolutional layers for the final predictions.
AttnCRNN also takes spectrograms as inputs and consists of two convolutional layers and two bi-directional LSTM layers with an attention module at the end.
The result shows that the TF-CRNN model with the Res-SE block achieves slightly better performance than the two previous works. 

\begin{figure}[t]
    \centering
    \includegraphics[width=\linewidth,keepaspectratio]{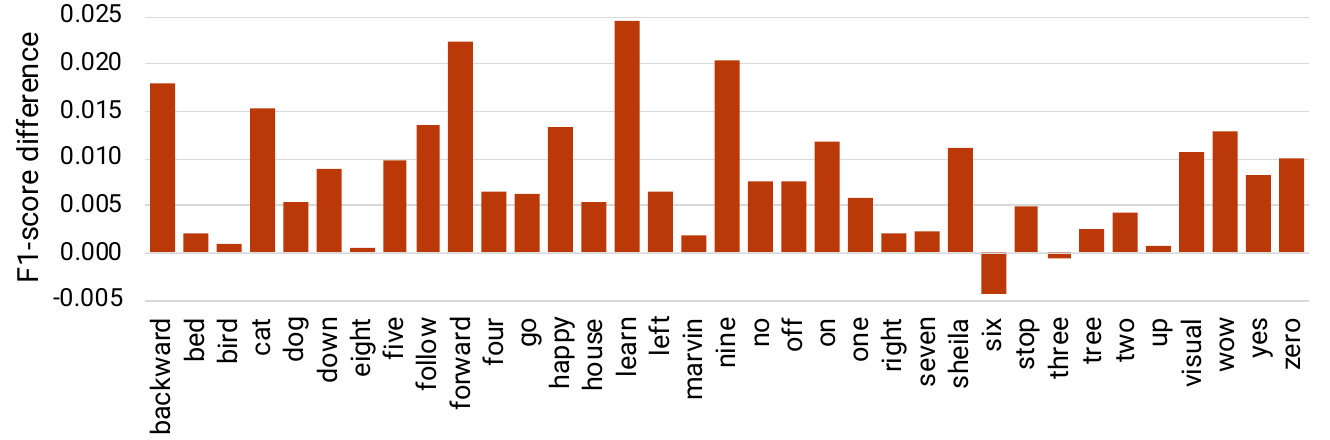}
    
    \includegraphics[width=\linewidth,keepaspectratio]{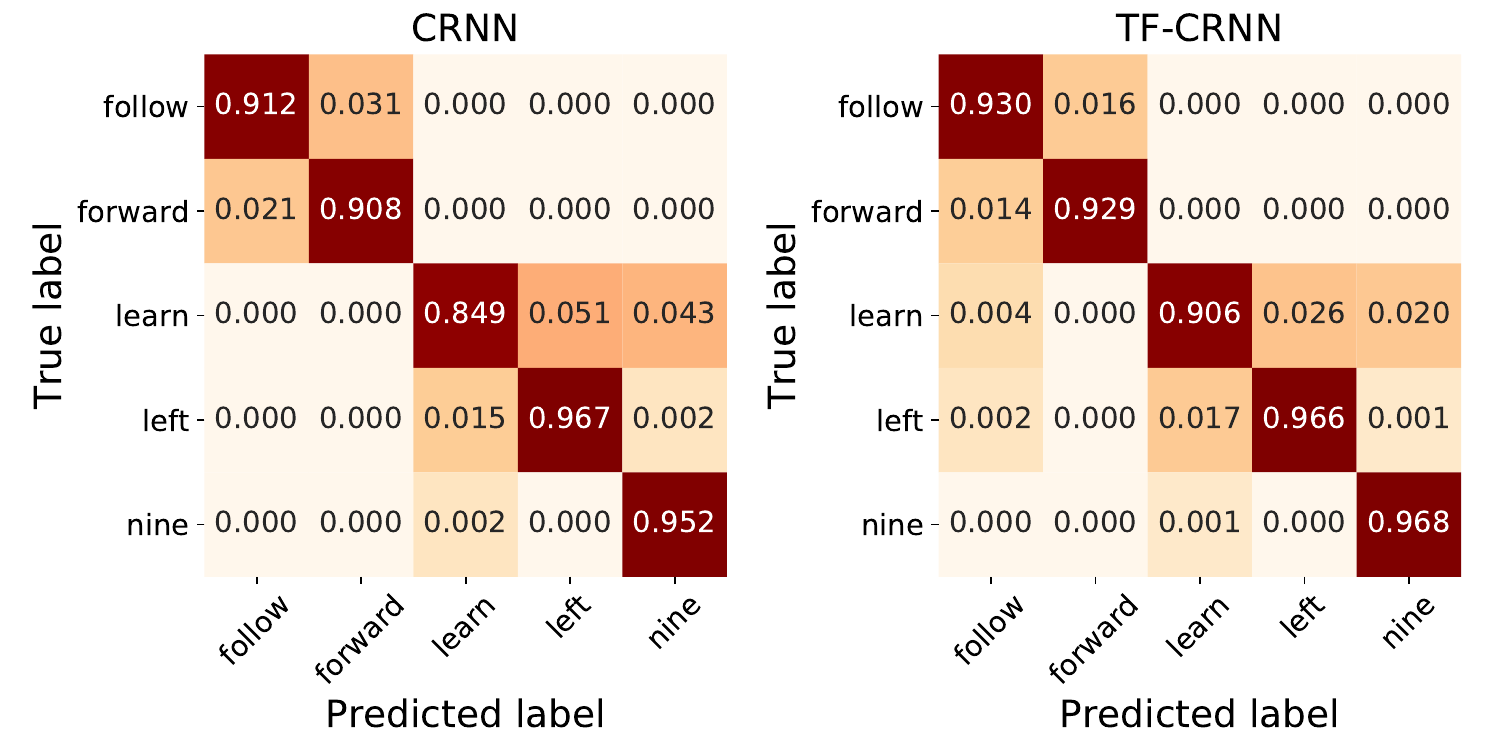}
    \caption{
    Performance differences between a CRNN and a TF-CRNN in each command averaging 3 experimental result for each model.
    Top: F1-score differences in all 35 commands.
    Bottom: Confusion matrices of each model for 5 example commands.
    }
    \label{fig:conf}
\end{figure}

\section{Analysis}

\subsection{Failure Analysis}
We investigate the effects of the temporal feedback further by conducting a failure analysis of misclassified commands. The top of \figurename~\ref{fig:conf} shows F1-score differences between CRNNs and TF-CRNNs for each command. The positive value indicates that TF-CRNNs outperform CRNNs for each command. We can observe that the performance differences are prominent for ``learn", ``forward", ``nine", ``backward", ``cat'' and so on. To analyze the errors and differences in details, we visualize the confusion matrices for the two models in the bottom of \figurename~\ref{fig:conf}. We selected top-3 commands (``forward", ``learn", and ``nine") and their confusing commands (``follow" and ``left"). We normalized the matrices by the total numbers of each true label, which displays the diagonal values as recall scores for each command. The models tend to misclassify a group of words with similar pronunciations such as ``learn" as ``left" or ``nine", and ``follow" as ``forward". However, these errors decrease by half in TF-CRNNs.

\begin{figure}[t]
    \centering
    \includegraphics[width=\linewidth,keepaspectratio]{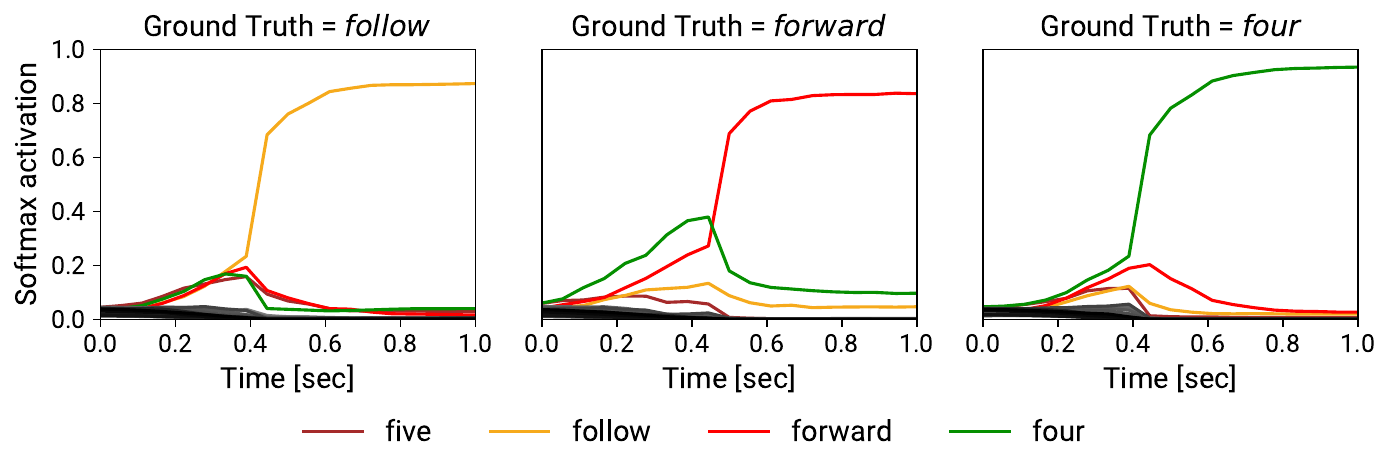}
    \caption{
      Changes in prediction probability over time for each command.
      Each plot is aggregated over samples of which ground truth are corresponding commands.
      Only the curves of the four example confusing commands are colored and the other commands are grayscale.
    }
    \label{fig:predcurve}
\end{figure}

Furthermore, we also visualize how prediction probabilities of confusing commands change over time in \figurename~\ref{fig:predcurve} using the TF-CRNN with the basic block and many-to-many setup on the test set.
Until arriving in the middle part of sequences, the probabilities of ground truth and confusing commands increase together but the probabilities of confusing commands decrease after the middle part.
As the figure shows, the CRNN can predict correctly from the middle of sequences, which might indicates many-to-many setup is more appropriate than many-to-one for speech command recognition.
Also, only a few command candidates are left already at the early stage of sequences so giving feedback from high-level representation to low-level feature extractor might affect the extractor to generate discriminative features focusing on a few candidate candidates.

\begin{figure}[t]
    \centering
    \includegraphics[width=\linewidth,keepaspectratio]{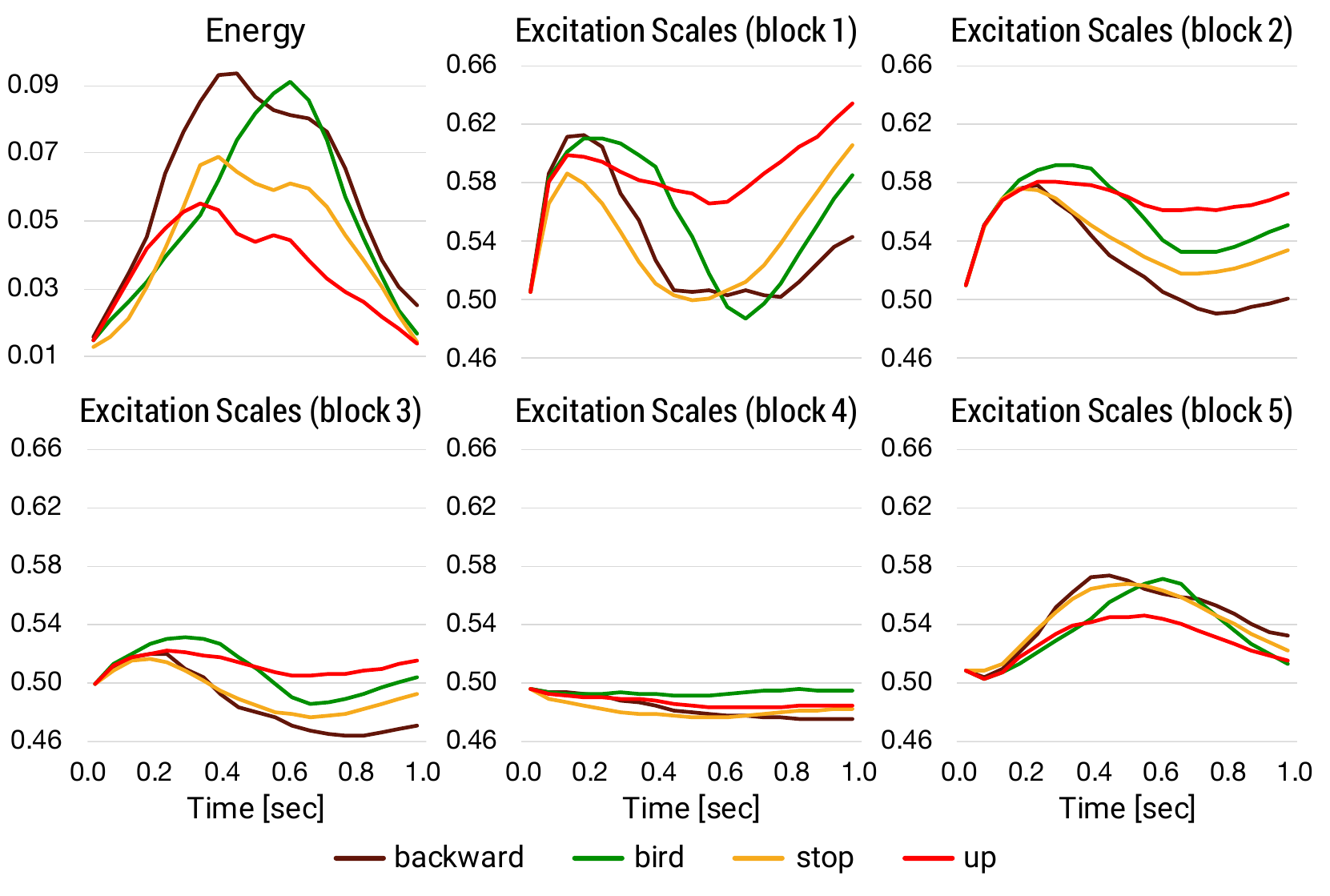}
    \caption{
    Amplitude envelope of audios (top left) and excitations averaged across all channels in each blocks (others). The values are averaged over each class.
    Note that the y-axes have the same range except for the amplitude envelope.
    }
    \label{fig:analysis}
\end{figure}

\subsection{Analysis of Temporal Feedbacks}
\label{sec:analysis_tf}
The role of temporal feedback in TF-CRNN is adjusting the strength of channel-wise activations using accumulated information from the upper level and, by doing so, increasing the representational power of the model. To understand the behaviors of TF-CRNN better, we computed statistics of the excitation scales over time from the temporal feedback in speech command recognition. We performed the analysis on a trained TF-CRNN with the many-to-many setup using the test set in speech command recognition. \figurename~\ref{fig:analysis} shows the summarized temporal excitation scales of four different commands (``backward'', ``bird'', ``stop'', and ``up'') from the first to the last block. Comparing them to root mean square (RMS) energy curves of input waveforms, we can observe that the excitation scales in the first block have opposite trends to the energy curves. When the energy levels increase, the scales become smaller and in turn they suppress the sensitivity of feature activations. When the energy levels decrease, the scales amplify the sensitivity. This behavior was analogous to the operation in the outer hair cells \cite{Lyon:2017}. The SENets also have similar patterns in the excitation scales~\cite{kim2019comparison} but the difference is that the temporal feedbacks in TF-CRNN is from the upper layer and they are used to normalize input signals in the next time step. As the layer goes up, the amplitude of the scales becomes attenuated. In the last layer, however, the trends are flipped (closer to the energy curves) and become more discriminant for the four classes. This is also similar to the trend in the analysis of SENets~\cite{hu2018senet,kim2019comparison}.

\section{Conclusion}
\label{sec:conclusion}
We proposed TF-CRNN, a novel architecture of neural networks inspired by the efferent connections in the human auditory processing. TF-CRNN performs channel-wise scaling in convolutional blocks by taking temporal feedbacks from the RNN module and controls the gain of channel-wise feature activations. We evaluated our models on the speech command recognition task and showed that the proposed model achieves superior performance. Finally, we conducted a failure analysis to examine the improvement in detail and visualized the excitation scales to better understand the behaviour of temporal feedback. For future work, we plan to extend TF-CRNN to other audio classification tasks such as music tagging or acoustic scene classification.



\bibliographystyle{IEEEbib}
\bibliography{main}

\end{document}